\documentclass[amstex]{article}
\usepackage{graphicx}
\usepackage{dcolumn}
\usepackage{amsmath}
\usepackage{amssymb}
\usepackage{amsfonts}
\usepackage{amscd}\newtheorem{thm}{Theorem}[section]
\newtheorem{lem}{Lemma}[section]
\newtheorem{prop}{Proposition}[section]

\begin{document}
\title{Decoherence free algebra}
\author{Yoshiko Ogata}
%\address{Department of Physics, Graduate School of Science, University of Tokyo,Hongo,7-3-1,Bunkyo-ku, Tokyo 113-0033, Japan}
%\email{ogata@monet.phys.s.u-tokyo.ac.jp}
\date{}
\maketitle
\begin{abstract}
We consider the decoherence free subalgebra which satisfies the minimal
condition introduced by Alicki \cite{alicki}.
We show the manifest form of it and relate the subalgebra
with the Kraus representation.
The arguments also provides a new proof for generalized 
L\"{u}ders theorem \cite{luders}.
\end{abstract}
\section{Introduction}
Decoherence is a non-unitary dynamics that is a consequence of system-
environment coupling.
As a way to protect a quantum system against decoherence,
a subalgebra whose dynamics is implementable by unitary
operator has attracted considerable interests;
that is, the decoherence free subalgebra.

Let $M_n$ be a $C^*$-algebra of all $n\times n$ matrices and
$\phi$ a unital trace preserving completely positive map
from $M_n$ into itself.
It is known that $\phi$ satisfies the Kadison inequality:
\begin{align}
\phi(x^*x)\ge\phi(x)^*\phi(x)\quad {\rm for\;all\;} x\in M_n.
\label{kadison}
\end{align}
The equality holds for all $x\in M_n$ if and only if $\phi$ is
given by a unitary operator $U$ as $\phi(x)=U^*xU$.
So $\phi(x^*x)-\phi(x)^*\phi(x)$ can be interpreted
as a dissipation function.
Under these considerations, Alicki introduced 
the {\it decoherence free subalgebra} $N_\phi$ of $\phi$ as \cite{alicki}
\begin{align*}
N_\phi=\{x_1+ix_2\in
M_n;x_i=x_i^*,\phi(x_i^2)=\phi(x_i)^2,\;i=1,2\}.
\end{align*}
It is a $C^*$-subalgebra of $M_n$. 
Suppose that $\phi$ can be joined to the identity map ${\rm id}$,
in the sense that there exists a continuous map from
interval $[0,1]$ to completely positive map
$t\to \phi_t$ which satisfies $\phi_0={\rm id}$,
and $\phi_1=\phi$.
Then there exists a unitary operator $U_\phi$ in $M_n$
such that
\begin{align*}
\phi(x)=U_\phi^*xU_\phi\;\;\;{\rm for\;all}\;x\in{N}_\phi.
\end{align*}
Hence it seems suitable to call $N_\phi$
{\it decoherence free algebra}.
However, how can we find the manifest
form of this subalgebra?
In particular, how is it related to the Kraus representation
$\phi(x)
=\sum_{i=1}^{k}A_i^*xA_i$?
In this paper, we consider this problem
and show the following proposition:
\begin{prop}\label{df}
The decoherence free subalgebra
$N_\phi$ is equal to the commutant of the algebra generated by $A_iA_j^*$;
\begin{align*}
N_\phi=\{A_iA_j^*,for\; all\;1\le i,j\le k\}^{'},
\end{align*}
which is independent of the choice of the Kraus representation.
\end{prop}
Here and hereafter, we denote the commutant of ${\cal C}$ by
${\cal C}'\equiv\{x\in M_n;xy=yx\;for\; all\;y\in{\cal C}\}$.
The arguments also provides a new proof for generalized 
L\"{u}ders theorem \cite{luders}.

\section{The decoherence free algebra}
Let us first confirm the notations.
We denote by $B({\cal H})$ the set of all bounded operators
on Hilbert space $\cal H$.
A subset $S$ of Hilbert space $\cal H$ is 
said to be total if it spans ${\cal H}$.
We denote by ${\mathfrak h}_m$ the $m$-dimensional Hilbert space,
and denote the orthonormal basis of it as
$\{{\rm e}_i;1\le i\le m\}$.
A linear map $\phi$ from $B({\cal H}_2)$ to
$B({\cal H}_1)$ is called a unital completely positive map
if it satisfies (i)$\phi(1)=1$
and (ii)for every $n=1,2,\cdots$ the correspondence
$(X_{ij})\to(\phi(X_{ij}))\;1\le i,j\le n$ from
$B({\cal H}_2\otimes{\mathbb C}^n)$ to   
$B({\cal H}_1\otimes{\mathbb C}^n)$ preserves positivity.

Before going into the proof, we introduce two famous Theorems.
The first one is the Theorem of Choi \cite{choi};
\begin{thm}
Suppose that $\phi$ is a unital completely positive map 
from $B({\cal H}_2)$
to $B({\cal H}_1)$.
If $\phi(a^*a)=\phi(a)^*\phi(a)$ and $\phi(aa^*)=\phi(a)\phi(a)^*$
for some $a\in B({\cal H}_2)$,
then for all $b\in B({\cal H}_2)$, we have
$\phi(ba)=\phi(b)\phi(a),\phi(ab)=\phi(a)\phi(b)$.
\end{thm}
The second one is the Theorem of Steinspring \cite{stinespring};
\begin{thm}
Let ${\cal H}_1$ and ${\cal H}_2$ be finite
dimensional Hilbert spaces
and let $\phi$ be a unital completely positive map 
from a $B({\cal H}_2)$ to $B({\cal H}_1)$.
Then there exists a Hilbert space ${\cal K}$ and
isometry $V$ from ${\cal H}_1$ to ${\cal H}_2\otimes {\cal K}$
such that the following holds:
\begin{description}
\item [(a)]
$\phi(x)=V^*(x\otimes 1)V\quad for\;all\;x\in B({\cal H}_2)$
\item[(b)]
$\{(x\otimes 1)Vu\;\vert\; x\in B({\cal H}_2),u\in{\cal H}_1\}$
is total in ${\cal H}_2\otimes {\cal K}$.
\item[(c)]
If ${\cal K}'$ is another Hilbert space and
$V'$ is an isometry 
from ${\cal H}_1$ to ${\cal H}_2\otimes {\cal K}'$
such that (a) and (b) hold, then there exists a unitary $W$ from
${\cal H}_2\otimes {\cal K}$ to ${\cal H}_2\otimes {\cal K}'$
such that $W(x\otimes 1_{\cal K})V=(x\otimes 1_{{\cal K}'})V'$.
\end{description}
\end{thm}
What we consider is a unital completely positive map from
$B({\cal H})$ to $B({\cal H})$,
where ${\cal H}$ is $n$-dimensional Hilbert space and $B({\cal H})$
is equal to $M_n$.
Now let us proceed to the proof of the proposition.\\\\
{\bf Proof of Proposition\ref{df}}:
Let $A_i\in B({\cal H}),1\le i\le m$ be Kraus operators which give the 
Kraus representation of $\phi$ as
\begin{align*}
\phi(a)=\sum_{1=1}^{m}A_i^*aA_i.
\end{align*}
Define a linear map $V$ from $\cal H$ to ${\cal H}\otimes{\mathfrak h}_m$
by
\begin{align*}
Vu=\sum_{i=1}^{m}A_iu\otimes{\rm e}_i.
\end{align*}
The adjoint of $V$ satisfies
\begin{align*}
V^*\left(\sum_{i=1}^{m}\psi_i\otimes{\rm e}_i\right)
=\sum_{i=1}^{m}A_i^*\psi_i,
\end{align*}
and 
\begin{align*}
V^*(a\otimes 1)V=\sum_{1=1}^{m}A_i^*aA_i=\phi(a).
\end{align*}
Therefore, the condition (a) of the 
Steinspring's Theorem holds for
the pair $(K\equiv{\mathfrak h}_m,V)$.

First we consider the case that the condition (b) is also satisfied.
Note that
by the definition of $N_\phi$, 
and the Choi's Theorem,
any element $a$ of $N_\phi$ satisfies
$\phi(a^*a)=\phi(a)^*\phi(a)$.
Then again by Choi's Theorem, we have
\begin{align}
\phi(ba)=\phi(b)\phi(a),\phi(ab)=\phi(a)\phi(b)
\label{choi}
\end{align}
for all $a\in N_\phi$ and all $b\in B({\cal H})$.
The relation (\ref{choi}) is represented in 
the Steinspring representation as
\begin{align*}
VV^*(a\otimes 1_K)VV^*(b\otimes 1_K)Vu
=VV^*(a\otimes 1_K)(b\otimes 1_K)Vu
\end{align*}
for all $u\in{\cal H}$, $a\in N_\phi$ and $b\in B({\cal H})$.
As $\{(b\otimes 1_K)Vu\vert b\in B({\cal H}),u\in{\cal H}\}$
is total in ${\cal H}\otimes K$
by condition (b), this implies
\begin{align*}
VV^*(a\otimes 1_K)VV^*=VV^*(a\otimes 1_K),
\end{align*}
on ${\cal H}\otimes K$.
The decoherence free algebra $N_\phi$ is involutive.
So we also have
\begin{align*}
VV^*(a^*\otimes 1_K)VV^*=VV^*(a^*\otimes 1_K).
\end{align*}
for all $a\in N_\phi$.
Taking the adjoint of this, we have
\begin{align*}
VV^*(a\otimes 1_K)VV^*=(a\otimes 1_K)VV^*.
\end{align*}
Combining these, we get
\begin{align*}
(a\otimes 1_K)VV^*=VV^*(a\otimes 1_K),
\end{align*}
for all $a\in N_\phi$.
As
\begin{align*}
VV^*=\sum_{i,j=1}^{m}A_iA_j^*\otimes
\vert{\rm e}_i\rangle\langle{\rm e}_j\vert,
\end{align*}
we obtain
\begin{align}
aA_iA_j^*=A_iA_j^*a\;\;\;1\le i,j\le m
\label{com}
\end{align}
for all $a$ in $N_\phi$.
Conversely, suppose that a self-adjoint element $a$ in $B({\cal H})$ satisfies
(\ref{com}).
Then it follows that
\begin{align*}
\phi(a)^2=\sum_{i,j=1}^{m}A_i^*aA_iA_j^*aA_j
=\sum_{i,j=1}^{m}A_i^*A_iA_j^*a^2A_j\\
=\phi(1)\phi(a^2)=\phi(a^2).
\end{align*}
Here we used the fact that $\phi$ is unital.
Hence $N_\phi$ is equal to the commutant
\begin{align*}
N_\phi=
\{A_iA_j^*\;\;1\le i,j\le k\}^{'}.
\end{align*}

Second we consider the case that condition (b) of the Steinspring's Theorem
fails.
Let $\tilde{P}$ be a projection operator in ${\cal H}\otimes {\mathfrak h}_m$
onto the subspace spanned by
$\{(x\otimes 1)Vu\;\vert\; x\in B({\cal H}),u\in{\cal H}\}$.
Note that $\tilde P$ satisfies 
\begin{align*}
{\tilde P}(x\otimes 1)(y\otimes 1)Vu
=(x\otimes 1)(y\otimes 1)Vu
=(x\otimes 1){\tilde P}(y\otimes 1)Vu.
\end{align*}
This implies
\begin{align*}
{\tilde P}(x\otimes 1){\tilde P}=(x\otimes 1){\tilde P},\quad
{\tilde P}(x^*\otimes 1){\tilde P}=(x^*\otimes 1){\tilde P},\quad x\in 
B({\cal H}).
\end{align*}
Taking the adjoint of the second equation, we obtain
\begin{align*}
(x\otimes 1){\tilde P}={\tilde P}(x\otimes 1),\quad x\in B({\cal H}).
\end{align*}
By Theorem of 5.9 of \cite{takesaki}, all the bounded operator
which commutes with $B({\cal H})\otimes 1$ belongs to
$1\otimes M_m$.
So ${\tilde P}$ is of the form ${\tilde P}=1\otimes P$,
where $P$ is a projection in ${\mathfrak h}_m$.
If (b) is not satisfied, $P\neq 1$.
In this case, we define a new orthogonal basis 
$f_j$ of ${\mathfrak h}_j$
so that there exists $1\le l< m$, such that
\begin{align}
&Pf_j=f_j\quad {\rm for}\quad 1\le j\le l\nonumber\\
&Pf_j=0\quad {\rm for}\quad l+1\le j\le m.
\label{fdef}
\end{align}
The orthonormal basis $\{e_i\}$ and $\{f_j\}$ are connected by
a unitary transformation, as
\begin{align*}
e_i=\sum_{j=1}^{m}v_{i,j}f_j,
\end{align*}
Define bounded operators $B_j\;1\le j\le m$ as
\begin{align*}
B_j=\sum_{i=1}^{m}v_{i,j}A_i.
\end{align*}
Using these relations, we obtain
\begin{align*}
Vu=\sum_{i=1}^{m}A_iu\otimes e_i
=\sum_{j=1}^{m}B_ju\otimes f_j.
\end{align*}
From $(1\otimes (1-P))Vu=0$ and (\ref{fdef}), 
we have $B_j=0$ for $l+1\le j\le m$.
We then see that condition (b) is satisfied
if and only if $A_i$ are linear independent.
Now as 
$\{(x\otimes 1)Vu\;\vert\; x\in B({\cal H}),u\in{\cal H}\}$
is total in 
${\cal H}\otimes P{\mathfrak h}_m={\cal H}\otimes {\mathfrak h}_l$,
the pair $({\mathfrak h}_l,V)$ satisfies conditions
(a),(b) of the Stinespring's Theorem.
So by the preceding arguments, we have
\begin{align*}
N_\phi=
\{B_iB_j^*\;\;1\le i,j\le l\}^{'}.
\end{align*}
On the other hands, as $v_{i,j}$ satisfies
\begin{align*}
\sum_{j=1}^m\overline{v_{i,j}}v_{k,j}=\delta_{i,k},
\end{align*}
we have
\begin{align*}
\sum_{j=1}^m \overline{v_{i,j}}B_j
=\sum_{k,j=1}^m \overline{v_{i,j}}v_{k,j}A_k=A_i.
\end{align*}
Hence we have
\begin{align*}
\{B_iB_j^*\;\;1\le i,j\le l\}^{'}
\supset \{A_iA_j^*\;\;1\le i,j\le m\}^{'},\\
\end{align*}
and
\begin{align*}
\{B_iB_j^*\;\;1\le i,j\le l\}^{'}
\subset \{A_iA_j^*\;\;1\le i,j\le m\}^{'},
\end{align*}
Therefore, we obtain
\begin{align*}
N_\phi=
\{B_iB_j^*\;\;1\le i,j\le l\}^{'}
=\{A_iA_j^*\;\;1\le i,j\le m\}^{'},
\end{align*}
which completes the proof.
$\square$\\\\
Let us consider the relation between decoherence free algebra $N_\phi$
and the fixed points $M_\phi$ of $\phi$ defined by
\begin{align*}
M_\phi=
\{x\in M_n\vert \phi(x)=x\}.
\end{align*}
We have the following Lemma:
\begin{lem}
Let $\phi$ be a unital trace preserving completely positive map
from $M_n$ into itself.Then $N_\phi$ includes $M_\phi$.
\end{lem}
{\it Proof}\\
Any $x\in M_\phi$ is decomposed into 
$x=x_1+ix_2$, where $x_1,x_2$ are self adjoint elements of $M_\phi$.
So it suffices to show that any self-adjoint element $x$ of $M_\phi$
satisfies $\phi(x^2)=\phi(x)^2$.
By the inequality (\ref{kadison}), we have
\begin{align*}
\phi(x^2)-x^2\ge 0.
\end{align*}
On the other hands, we have
\begin{align*}
{\rm Tr}\left(\phi(x^2)-x^2\right)=0,
\end{align*}
because $\phi$ is trace preserving.
As trace is faithful, we have
\begin{align*}
\phi(x^2)=x^2=\phi(x)^2,
\end{align*}
which completes the proof.
$\square$

Let $\cal A$ be a $C^*$-algebra generated by $A_i$, and
$\cal B$, a $C^*$-algebra generated by $A_iA_j^*$.
Clearly, ${\cal B}\subset {\cal A}$ and 
the following inclusion relation holds:
\begin{align*}
{\cal A}'
\subset M_\phi\subset N_\phi
={\cal B}'
\end{align*}
If $A_i$ are positive, then $A_i=\sqrt{A_iA_i^*}\in {\cal B}$,
i.e., all the generators of ${\cal A}$ is included in ${\cal B}$.
So we have
${\cal A}={\cal B}$, which implies ${\cal A}'={\cal B}'$ .
In this case, we have 
\begin{align*}
{\cal A}'= M_\phi=N_\phi={\cal B}'.
\end{align*}
That is, all the decoherence free elements are
fixed points if $A_i$ are positive. 
This is a new proof of a generalized L\"{u}ders theorem \cite{luders},
\cite{fixed}.
%\acknowledgements{The author would like to thank Prof.M.Wadati for valuable com%ments and critical reading of the manuscript,
%and also thank Dr.A.Miyake for helpful discussions.}

\end{document}